\documentclass{article}

\usepackage[english]{babel}                                                                                           

\usepackage[letterpaper,top=2cm,bottom=2cm,left=3cm,right=3cm,marginparwidth=1.75cm]{geometry}

\usepackage{amsmath}
\usepackage{graphicx}
\usepackage[colorlinks=true, allcolors=blue]{hyperref}
\usepackage{amsthm}
\usepackage{tikz}
\usepackage{centernot}
\usepackage{mathtools}
\usepackage{cite}
\usepackage{ stmaryrd }
\usetikzlibrary{matrix,arrows.meta}
\usepackage{amsmath}
\usepackage{tikz}
\usetikzlibrary{matrix,arrows,decorations.pathmorphing}
\usetikzlibrary{shapes,arrows,positioning}
\usepackage{graphicx}
\usepackage{amsmath}
\usepackage{amssymb}
\usepackage{latexsym}
\usepackage{amssymb,amsmath}
\usepackage{tikz-cd}
\usetikzlibrary{arrows}
\usepackage{amsfonts}
\usepackage{amssymb}
\usepackage{graphicx}
\usepackage{fancyhdr}
\usepackage{fancyvrb}
\usepackage{fancybox}
\usepackage{float}
\usepackage{geometry}
\usepackage{minitoc}
\usepackage{indentfirst}
\usepackage{multicol, color}
\usepackage[outerbars]{changebar}
\usepackage{pifont,textcomp}
\usepackage{amsfonts}
\usepackage{amsmath}
\usepackage{amscd}
\usepackage{array}
\usepackage{multirow}
\usepackage{mathrsfs}
\usepackage{amssymb}
\usepackage{amsthm}
\usepackage{amsmath}
\usepackage{subfig}
\usepackage{graphicx}
\usepackage{braket}
\usepackage[colorlinks=true, allcolors=blue]{hyperref}

 \title{Squeezed Coherent States in Supersymmetric Quantum Mechanics with Position-Dependent Mass}
\author{  Daniel Sabi Takou$^{1,2}$, 	Amidou Boukari$^{2}$, 	Assimiou Yarou Mora$^{2}$  and  Gabriel Y. H. Avossevou$^{2}$
 	\space\\\\
 	$^{1}$Ecole Polytechnique d'Abomey Calavi (EPAC-UAC),\\
	Universit\'e d'Abomey-Calavi (UAC), B\'enin\\
	$^{2}$ Unit\'e de Recherche en Physique Th\'eorique (URPT),\\
	Institut de Math\'ematiques et de Sciences  Physiques (IMSP),\\
		01 B.P. 613 Porto-Novo, Rep. du B\'enin\\\\		 
 sabitakoudaniel11@gmail.com$^{1,2}$,  amidouboukari12@gmail.com$^{2}$,\\  assimiouyaroumora@gmail.com $^{3}$ and gabriel.avossevou@imsp-uac.org$^{2}$ }

\date{}
\date{}

\begin{document}
\maketitle
\begin{abstract}
In this paper, we construct and analyze a class of squeezed coherent states within the framework of supersymmetric quantum mechanics (SUSYQM) involving a position-dependent mass (PDM). Using a deformed algebraic structure, we generalize the creation and annihilation operators to accommodate spatially varying mass profiles. The resulting states exhibit non-classical features, such as squeezing, coherence, and modified uncertainty relations, strongly influenced by both the deformation parameters and the mass function. We explore their physical properties through expectation values, variances, and probability densities. This work provides a pathway toward extending coherent state theory to more complex quantum systems with geometrical and algebraic richness.
\end{abstract}


\maketitle
\section{Introduction} \label{sec1}

Coherent and squeezed states play a fundamental role in various domains of physics, particularly quantum optics and quantum information theory\cite{1,2,3}. Originally introduced by Schrödinger in 1926 and Kennard in 1927\cite{4,5}, these states remained largely overlooked until the 1960s, when seminal works by Glauber, Klauder, and Nieto reignited interest in their theoretical and practical significance\cite{551,552,553}. Squeezed states, in particular, are nonclassical quantum states that exhibit reduced quantum noise, making them highly valuable for optical communication, precision measurements, and the detection of gravitational waves---as in the case of LIGO\cite{554,555,556,557}. They are also key resources in continuous-variable quantum information processing, including quantum computing, dense coding, and quantum cryptography, and can be implemented in physical platforms such as graded semiconductors~\cite{Bastard1988,Peter2009}, quantum dots, photonic crystals, and cavity optomechanical systems for quantum communication and precision sensing~\cite{30,Ullah2020}.
 More recently, the construction and analysis of squeezed states for systems with infinite discrete spectra have attracted increasing attention as generalized extensions of coherent states  (see, for
 example\cite{13aa1,13aa2,13aa3,13a1,13a4}).

Quantum mechanics with position-dependent mass (PDM) has emerged as a powerful framework for modeling a wide range of physical systems with non-uniform spatial characteristics, such as semiconductor heterostructures, graded crystals, and quantum wells \cite{6a,6b,6c,6d,10a1,6d1}. The modification of the kinetic energy operator due to the spatial dependence of the mass introduces new challenges and has stimulated the development of refined analytical and algebraic methods. Among these, supersymmetric quantum mechanics (SUSY QM) has proven particularly fruitful in the construction of exactly solvable models through factorization methods and intertwining techniques \cite{6dd1,6dd2}.

On the other hand, coherent and squeezed states, first introduced by Schrödinger in 1926 and later formalized by Glauber, Klauder, and others \cite{7,8,9}, have become essential tools in quantum optics and quantum information due to their ability to saturate uncertainty relations and describe non-classical light fields.These states have been widely generalized to encompass systems with either discrete or continuous spectra, algebraic deformations, and both constant and position-dependent mass backgrounds. Further extensions have also been developed for more complex scenarios, such as oscillators defined in noncommutative spaces \cite{dehv,dehv1,dehv2}.

In this work, we aim to construct a new class of generalized squeezed coherent states for a quantum system governed by a position-dependent mass and embedded within a supersymmetric algebraic structure. Inspired by the formalism developed by Gazeau and Klauder \cite{7}, and following techniques involving ladder operators and recurrence relations, we derive analytical expressions for the coherent-like states, ensuring that they fulfill essential physical criteria such as temporal stability, resolution of identity, and continuity in the labeling parameter \cite{8,9}.

Moreover, we devote special attention to the analysis of quantum statistical properties of these states. In particular, we compute their normalization factors, probability densities, and expectation values of relevant observables. The wavefunctions and energy spectrum are explicitly obtained and expressed in terms of orthogonal polynomials and generalized hypergeometric functions, revealing the deep interplay between the effective mass profile, deformation parameters, and the algebraic structure of the system.

 Our manuscript is organised as follows : In Section~\ref{sec2}, we review the basic formalism of position-dependent mass quantum mechanics. Section~\ref{sec3}  is devoted to the study of the  Supersymmetric Squeezed coherent states. In Section~\ref{sec4}, we investigate the properties of squeezed states and discuss numerical results. Finally, concluding remarks are presented in Section~\ref{sec5}.

\section{Position-Dependent Effective Mass Harmonic Oscillator and it Exact Solutions}\label{sec2}

A general expression for the Hermitian kinetic energy operator of a quantum particle with a position-dependent mass (PDM) \( m(\hat{x}) \) is given by \cite{10,10a1,10a2}
\begin{equation}
\hat{T} = \frac{1}{4} \left[ m^\alpha(\hat{x}) \hat{p} \, m^\beta(\hat{x}) \hat{p} \, m^\gamma(\hat{x}) + m^\gamma(\hat{x}) \hat{p} \, m^\beta(\hat{x}) \hat{p} \, m^\alpha(\hat{x}) \right],
\end{equation}
where the parameters \( \alpha \), \( \beta \), and \( \gamma \) satisfy the constraint \( \alpha + \beta + \gamma = -1 \). Different combinations of these parameters lead to various forms of the kinetic energy operator~\cite{11,11a,11b,11c,11d}.

In this study, we adopt the representation proposed by Mustafa and Mazharimousavi~\cite{11d}, which corresponds to the specific parameter choice \( \alpha = \gamma = -\frac{1}{4} \), \( \beta = -\frac{1}{2} \). Under this configuration, the kinetic energy operator becomes:
\begin{equation}
\hat{T} = \frac{1}{2} \frac{1}{m^{1/4}(\hat{x})} \, \hat{p} \left( \frac{1}{m^{1/2}(\hat{x})} \right) \hat{p} \, \frac{1}{m^{1/4}(\hat{x})}.
\end{equation}

Accordingly, the total Hamiltonian for the system is written as
\begin{equation}
\hat{H} = \hat{T} + V = \frac{1}{2} \frac{1}{m^{1/4}(\hat{x})} \, \hat{p} \left( \frac{1}{m^{1/2}(\hat{x})} \right) \hat{p} \, \frac{1}{m^{1/4}(\hat{x})} + V(\hat{x}),
\end{equation}
where \( V(\hat{x}) \) denotes the potential energy function.

Considering a harmonic oscillator potential defined by \( V(x) = \frac{1}{2} m_0 \omega^2 x^2 \), the corresponding time-independent Schrödinger equation reads:
\begin{equation}\label{a}
E \phi(x) = -\frac{\hbar^2}{2m_0} \sqrt[4]{\frac{m_0}{m(x)}} \frac{d}{dx} \sqrt{\frac{m_0}{m(x)}} \frac{d}{dx} \sqrt[4]{\frac{m_0}{m(x)}} \phi(x) + V(x) \phi(x),
\end{equation}
where \( m_0 \) represents the constant reference mass, \( E \) is the energy eigenvalue, and \( \phi(x) \) is the wavefunction belonging to the Hilbert space \( \mathcal{H} = \mathcal{L}^2(\mathbb{R}) \).

In this work, we consider the following mass distribution:
\begin{equation}\label{71}
m(x) = \frac{m_0}{(1 + \alpha x^2)^2},
\end{equation}
where \( \alpha \) is a deformation parameter constrained by \( 0 < \alpha < 1 \). This choice generalizes earlier mass profiles studied in~\cite{6a,6b,6c,6d,10a1}, and can be interpreted as an inverse squared length scale associated with local geometric features such as curvature or structural defects in the physical system.

To simplify Eq.~\eqref{a}, we apply the transformation \( \phi(x) = \sqrt[4]{\frac{m(x)}{m_0}} \psi(x) \), yielding:
\begin{equation}\label{schro}
E \psi(x) = -\frac{\hbar^2}{2 m_0} \left( \sqrt{\frac{m_0}{m(x)}} \frac{d}{dx} \right)^2 \psi(x) + \frac{1}{2} m_0 \omega^2 x^2 \psi(x).
\end{equation}

Using the explicit expression of \( m(x) \) in Eq.~\eqref{71}, the above equation becomes:
\begin{equation}\label{schro2}
E \psi(x) = -\frac{\hbar^2}{2 m_0} \left[ (1 + \alpha x^2) \frac{d}{dx} \right]^2 \psi(x) + \frac{1}{2} m_0 \omega^2 x^2 \psi(x).
\end{equation}
By introducing the coordinate transformation \( q = \arctan(x\sqrt{\alpha}) \), the infinite domain of the position variable \( x \in \mathbb{R} \) is mapped to a finite interval \( q \in \left( -\frac{\pi}{2}, \frac{\pi}{2} \right) \). This change significantly simplifies the position-dependent mass Schrödinger equation, which becomes more tractable in the new variable.

Assuming the ansatz \( \psi(q) = \cos^\lambda q \cdot f(\sin q) \), and requiring the absence of singular behavior at the boundaries, leads to a second-order differential equation that reduces to the Gegenbauer differential equation when a specific condition on the parameter \( \lambda \) is imposed:
\begin{eqnarray}
\lambda = \frac{1}{2} + \frac{1}{2} \sqrt{1 + 4\kappa^2}, \quad \kappa = \frac{m_0\omega}{\alpha\hbar}.
\end{eqnarray}
This constraint eliminates the singular term, yielding a well-defined eigenvalue problem with solutions expressible in terms of Gegenbauer polynomials \( C_n^{\lambda}(s) \), where \( s = \sin q \). The energy quantization condition is then given by:
\begin{eqnarray}
\varepsilon = \lambda + n(n + 2\lambda), \quad n \in \mathbb{N}.
\end{eqnarray}
Consequently, the energy spectrum becomes:
\begin{eqnarray}\label{en11}
E_n = \hbar\omega \left(n + \frac{1}{2} \right) \sqrt{1 + \frac{\alpha^2 \hbar^2}{4m_0^2 \omega^2}} + \frac{\alpha \hbar^2}{2m_0} \left(n^2 + 2n + \frac{1}{2} \right).
\end{eqnarray}
This reveals a non-trivial dependence on the deformation parameter \( \alpha \), illustrating how position-dependent mass modifies the energy levels compared to the standard harmonic oscillator\cite{14,15}. In the limit \( \alpha \to 0 \), the conventional energy levels are recovered.

The corresponding normalized wavefunctions take the form:
\begin{eqnarray}
\psi_n(q) = N \cos^\lambda q \cdot C_n^{\lambda}(\sin q),
\end{eqnarray}

where $ C_n^\lambda(q)$  are the Gegenbauer polynomials \cite{13} and $N$ the normalization constant given by :
\begin{eqnarray}
N = \sqrt{ \frac{n! (n + \lambda) \Gamma^2(\lambda)}{\pi 2^{1 - 2\lambda} \Gamma(n + 2\lambda)} }.
\end{eqnarray}
Transforming back to the \( x \)-representation using \( \cos q = \frac{1}{\sqrt{1 + \alpha x^2}} \) and \( \sin q = \frac{x\sqrt{\alpha}}{\sqrt{1 + \alpha x^2}} \), the wavefunction becomes:
\begin{eqnarray}
\phi_n(x) = \sqrt{ \frac{n! (n + \lambda) \Gamma^2(\lambda)}{\pi 2^{1 - 2\lambda} \Gamma(n + 2\lambda)} } \left( \frac{1}{\sqrt{1 + \alpha x^2}} \right)^{\lambda + 1} C_n^{\lambda} \left( \frac{x \sqrt{\alpha}}{\sqrt{1 + \alpha x^2}} \right).
\end{eqnarray}

This provides an explicit analytic solution for the eigenfunctions of the deformed harmonic oscillator with position-dependent mass.

\section{Supersymmetric Squeezed coherent states }\label{sec3}

The Hamiltonian operator given in Eq.~\eqref{schro2}, corresponding to a quadratic potential \( v(x) = \frac{1}{2} m_0 \omega^2 x^2 \), can be factorized in the form  
\begin{equation}
    \hat{H} = \hbar \omega_0\, \hat{a}_\alpha^\dagger \hat{a}_\alpha + E_0(\alpha),
\end{equation}
where the ladder operators \( \hat{a}_\alpha^\dagger \) and \( \hat{a}_\alpha \) are constructed within the supersymmetric framework for systems with position-dependent mass, as defined in~\cite{10a1}.
\begin{align}\label{AAd}
    \hat{a}_\alpha = \sqrt{\frac{m_0 \omega_0}{2\hbar}}\left( \hat{x} + \frac{\hbar \alpha}{2 m_0 \omega_0} + \frac{i}{m_0 \omega_0} \hat{\Pi}_\alpha \right), \;\;\;
    \hat{a}_\alpha^\dagger = \sqrt{\frac{m_0 \omega_0}{2\hbar}}\left( \hat{x} + \frac{\hbar \alpha}{2 m_0 \omega_0} - \frac{i}{m_0 \omega_0} \hat{\Pi}_\alpha \right),
\end{align}
where \( \hat{\Pi}_\alpha = (1 + \alpha \hat{x}^2) \hat{p} \) is the deformed momentum operator.

In supersymmetric framing, we used the fact that \( \hat{a}_\alpha \psi_0(x) = 0 \), indicating that the ground state is annihilated by the operator \( \hat{a}_\alpha \). Although \( \hat{a}_\alpha \) and \( \hat{a}_\alpha^\dagger \) factorize the Hamiltonian Eq.~\eqref{schro2} associated with the quadratic potential, they do not serve as conventional ladder operators, since they obey the deformed commutation relation:
$[\hat{a}_\alpha, \hat{a}_\alpha^\dagger] = \hat{1} + \alpha \hat{x}^2.$
To take into account this deformation, we introduce a modified number operator defined by: $\hat{n}_\alpha = \hat{a}_\alpha^\dagger \hat{a}_\alpha,$
whose expectation value on the \( n \)-th eigenstate can be obtained as follow. 

The generalized number operator
$\hat{n}_\alpha=\hat{a}_\alpha^\dagger\hat{a}_\alpha$,
with $\hat{a}_\alpha$ and $\hat{a}_\alpha^\dagger$ defined in \eqref{AAd}, can be written as
\begin{equation}
\hat{n}_\alpha
= \frac{m_0\omega_0}{2\hbar}(\hat{x}+c)^2
+ \frac{1}{2\hbar m_0\omega_0}\,\hat{\Pi}_\alpha^2
- \frac{1}{2}\big(1+\alpha\hat{x}^2\big),
\end{equation}
where $c=\frac{\hbar\alpha}{2m_0\omega_0}$ and 
$\hat{\Pi}_\alpha=(1+\alpha\hat{x}^2)\hat{p}$.  
Its expectation value in an arbitrary state $|\psi\rangle$ is
\begin{equation}
\langle\hat{n}_\alpha\rangle
= \frac{m_0\omega_0}{2\hbar}\langle(\hat{x}+c)^2\rangle
+ \frac{1}{2\hbar m_0\omega_0}\langle\hat{\Pi}_\alpha^2\rangle
- \frac{1}{2}\big(1+\alpha\langle\hat{x}^2\rangle\big).
\end{equation}
Physically, $\langle\hat{n}_\alpha\rangle$ measures the mean excitation number 
of the deformed oscillator defined by the position-dependent mass model.  
The shift $c$ introduces a static displacement of the quadrature $\hat{x}$, 
while $\hat{\Pi}_\alpha$ accounts for a deformation of the canonical momentum, 
modifying the balance between kinetic and potential contributions.  
In the limit $\alpha\to 0$, $\hat{n}_\alpha$ reduces to the standard harmonic 
oscillator number operator, and $\langle\hat{n}_\alpha\rangle$ counts the 
average quanta of excitation in the usual sense.

We define the squeezed coherent state \( \psi(z,\gamma,x) \) as the solution to the eigenvalue equation involving a linear combination of ladder operators, in accordance with the formalism presented in \cite{13a1,553}.
\begin{equation}\label{sq1}
    (\hat{a}_\alpha + \gamma \hat{a}_\alpha^\dagger) \ket{\psi(z,\gamma)} = z \ket{\psi(z,\gamma)}.
\end{equation}
This combination of the operators $\hat{a}_\alpha$ and $\hat{a}_\alpha^\dagger$ is governed by a squeeze parameter $\gamma$, while \( z \) is known as \emph{coherent parameter}. When the squeezing parameter vanishes, i.e., \( \gamma = 0 \), the resulting quantum states reduce to \emph{coherent states}, which are special solutions characterized by minimal uncertainty and classical-like behavior. To ensure that these states are physically meaningful, appropriate conditions must be imposed on the parameters involved, particularly to guarantee the normalizability of the states.

Squeezed coherent states (SCS) have been extensively studied in various algebraic frameworks, notably those based on the Lie algebras $\mathrm{su}(2)$ and $\mathrm{su}(1,1)$~\cite{13a2,13a21,13a3,13a4}. In addition, constructions have been extended to \emph{direct sums} of these algebras with the \emph{Heisenberg algebra} \( \mathrm{h}(2) \)~\cite{13a4}, allowing for a richer variety of quantum states.

In such algebraic settings, especially for \( \mathrm{su}(2) \) and \( \mathrm{su}(1,1) \), the function \( k(n) \) that appears in the recurrence relations or ladder operator actions is generally a \emph{quadratic function} of the quantum number \( n \). This reflects the non-equidistant spectrum structure and the non-linear characteristics of the underlying algebra, distinguishing these systems from the standard harmonic oscillator.

We expand this state \eqref{sq1} in the eigen basis of the deformed oscillator :
\begin{equation}\label{func1}
    \ket{\psi(z, \gamma)} = \frac{1}{\sqrt{\mathcal{N}(z, \gamma)}} \sum_{n=0}^{n_{\max}} \frac{Z(z,\gamma,n)}{\sqrt{\rho(n)}} \ket{\psi_n},
\end{equation}
where the normalization factor is computed as $\mathcal{N}(z, \gamma) = \sum_{n=0}^{n_{\max}} \frac{|Z(z, \gamma, n)|^2}{\rho(n)}$ and \( Z(z,\gamma,n) \) satisfies the recurrence relation:
\begin{equation}
    Z_{n+1}(z, \gamma) = z Z_n(z, \gamma) - \gamma k(n) Z_{n-1}(z, \gamma), \quad Z_0 = 1, \quad Z_1 = z.
\end{equation}

We define the weights \( \rho(n) \) using the generalized factorial:
\begin{equation}
    \rho(n) = \prod_{j=1}^{n} k(j).
\end{equation}
Let us recall the  energy eigenvalues defined in equation \eqref{en11} as
\begin{eqnarray}\label{en1}
    E_n=\hbar\omega \left(n+\frac{1}{2}\right)\sqrt{1+\frac{\alpha^2\hbar^2}{4m_0^2\omega^2}}+\frac{\alpha\hbar^2}{2m_0}\left(n^2+2n+\frac{1}{2}\right)\quad \mbox{and}\quad E_0=\frac{1}{2}\hbar\omega \sqrt{1+\frac{\alpha^2\hbar^2}{4m_0^2\omega^2}}+\frac{\alpha\hbar^2}{4m_0}.
\end{eqnarray}
For the system under consideration, the dimensionless form of the latter energy is given by
\begin{eqnarray}
    e_n&=&E_n-E_0=n\hbar\omega \sqrt{1+\frac{\alpha^2\hbar^2}{4m_0^2\omega^2}}+\frac{\alpha\hbar^2}{2m_0}\left(n^2+2n\right)= \left(\hbar\omega \sqrt{1+\frac{\alpha^2\hbar^2}{4m_0^2\omega^2}}+\frac{\alpha\hbar^2}{m_0}\right)n+\frac{\alpha\hbar^2}{2m_0}n^2\cr
    &=&an^2+bn=n(an+b),\label{q1}
\end{eqnarray}
where the constants $a$ and $b$ are given by
\begin{eqnarray}\label{q66}
    b=\hbar\omega \sqrt{1+\frac{\alpha^2\hbar^2}{4m_0^2\omega^2}}+\frac{\alpha\hbar^2}{m_0},\;\;\;a=\frac{\alpha\hbar^2}{2m_0}.
\end{eqnarray}
  The dimensionless form of the  energy \eqref{q1} is similar to the ones obtained in \cite{15,30} used to study the laser light propagation in a nonlinear Kerr medium.
The product of
these dimensionless energies $e_n$  represented by  $\rho_n$ is defined as
\begin{eqnarray}
\rho_n &=& \prod_{k=1}^n e_k, \quad\mbox{with}\quad e_i =  ak^2 + b k\cr
        &=& \prod_{k=1}^n k  \prod_{k=1}^n ( ak + b).\label{qq1}
\end{eqnarray}
With the following computations
\begin{eqnarray}
\prod_{k=1}^n k = n!=\Gamma(n+1), \quad \prod_{k=1}^n ( ak + b) = \prod_{k=1}^n a\left( k + \frac{b}{a}  \right)=\frac{a^n\Gamma\left( n + 1 + \frac{b}{a} \right)}{\Gamma\left( 1 + \frac{b}{a}  \right)}.
\end{eqnarray}
The equation \eqref{qq1} becomes 
\begin{eqnarray}
\rho_n = n!  a^n \frac{\Gamma\left( n + 1 + \frac{b}{a} \right)}{\Gamma\left( 1 + \frac{b}{a}  \right)}= \frac{a^n\Gamma\left( n + 1  \right)\Gamma\left( n + 1 + \frac{b}{a}  \right)}{\Gamma\left( 1 + \frac{b}{a}  \right)},\quad \rho_0=1.
\end{eqnarray}

To derive the analytical expression of the coefficients \( Z_n(z, \gamma) \) that define the squeezed coherent states in the deformed algebraic framework, we assume a factorized form:
\begin{equation}
    Z_n(z, \gamma) = \left( \frac{\gamma}{2} \right)^{n/2} f_n(\omega), \quad \text{with} \quad \omega = \frac{z}{\sqrt{2\gamma}}.
\end{equation}
Substituting this ansatz into the recurrence relation:
\begin{equation}\label{qq2}
    Z_{n+1} = z Z_n - \gamma\, n(an + b) Z_{n-1},
\end{equation}
leads to a simplified recurrence for \( f_n(\omega) \) of the form:
\begin{equation}
    f_{n+1} = 2\omega f_n - 2n(an + b) f_{n-1}, \quad f_0 = 1, \quad f_1 = \sqrt{2\gamma} \, \omega.
\end{equation}
This sequence admits an analytical solution expressed in terms of the Gauss hypergeometric function:
\begin{equation}
    f_n(\omega) = (2\omega)^n \cdot {}_2F_1\left(-\left\lfloor \frac{n}{2} \right\rfloor,\; -\left\lfloor \frac{n-1}{2} \right\rfloor;\; -b;\; 1 \right),
\end{equation}
where \( {}_2F_1(a,b;c;z) \) is the Gaussian hypergeometric function.

 The general solution of the recurrence relation \eqref{qq2} is obtained in terms of the Gauss hypergeometric function ${}_2F_1$ as follows
\begin{equation}
Z(z, \gamma,n) =(-1)^n \left( \gamma \right)^{n/2} a^{n/2}\frac{\Gamma(n+1+\frac{b}{a})}{\Gamma(1+\frac{b}{a})} \cdot {}_2F_1\left( -n,\; - \frac{1}{2}+\frac{b}{2a}-\frac{z}{2\sqrt{\gamma a}} ;\;1+\frac{b}{a};\; 2 \right).
\end{equation}

We obtain the explicit form of the squeezed states of position-dependent effective mass harmonic oscillator \eqref{func1} in terms of the hypergeometric functions as follows :
\begin{equation}\label{qq3}
\ket{\psi(z, \gamma)} = \frac{1}{\sqrt{\mathcal{N}(z, \gamma)}}
\sum_{n=0}^{n_{\max}} 
\frac{(\gamma a)^{n/2}}{\sqrt{n!}} \cdot 
\sqrt{\frac{\Gamma\left(n+1+\frac{b}{a}\right)}{\Gamma\left(1+\frac{b}{a}\right)}} \cdot
{}_2F_1\left(-n,\; -\frac{1}{2} + \frac{b}{2a} - \frac{z}{2\sqrt{a\gamma}};\; 1+\frac{b}{a};\; 2\right)
\ket{\psi_n}
\end{equation}

where

\begin{equation}\label{no1}
\mathcal{N}(z, \gamma) = 
\sum_{n=0}^{n_{\max}} 
\frac{(\gamma a)^n}{n!} \cdot 
\frac{\Gamma\left(n+1+\frac{b}{a}\right)}{\Gamma\left(1+\frac{b}{a}\right)} \cdot
\left|{}_2F_1\left(-n,\; -\frac{1}{2} + \frac{b}{2a} - \frac{z}{2\sqrt{a\gamma}};\; 1+\frac{b}{a};\; 2\right)\right|^2
\end{equation}

The time evolution of this squeezed coherent states \eqref{qq3} is given by

\begin{equation}\label{qq33}
\ket{\psi(z, \gamma,t)} = \frac{1}{\sqrt{\mathcal{N}(z, \gamma)}}
\sum_{n=0}^{n_{\max}} 
\frac{(\gamma a)^{n/2}}{\sqrt{n!}} \cdot 
\sqrt{\frac{\Gamma\left(n+1+\frac{b}{a}\right)}{\Gamma\left(1+\frac{b}{a}\right)}} \cdot
{}_2F_1\left(-n,\; -\frac{1}{2} + \frac{b}{2a} - \frac{z}{2\sqrt{a\gamma}};\; 1+\frac{b}{a};\; 2\right)e^{-\frac{iE_n }{\hbar}t}
\ket{\psi_n}
\end{equation}

We are interested in analyzing the special case of \eqref{q66} where the deformation parameter \( \alpha \to 0 \), which corresponds to the standard quantum harmonic oscillator with constant mass, corresponds to :

\begin{eqnarray}
    a = \frac{\alpha \hbar^2}{2m_0} = 0,\;\;\;
    b = \hbar\omega \sqrt{1+\frac{\alpha^2\hbar^2}{4m_0^2\omega^2}} + \frac{\alpha\hbar^2}{m_0} 
      &\xrightarrow{\alpha = 0} \hbar\omega \sqrt{1} + 0 = \hbar\omega.
\end{eqnarray}

In the undeformed limit \( \alpha = 0 \), we obtain the simplified and physically interpretable results:
\begin{equation}
    a = 0, \quad b = \hbar\omega.
\end{equation}

This implies that:
\begin{itemize}
    \item The deformation parameter vanishes, and the mass becomes position-independent;
    \item The parameter \( b \) reduces to the fundamental energy quantum \( \hbar\omega \), characteristic of the standard harmonic oscillator;
    \item The wavefunctions and the associated squeezed coherent states simplify accordingly, often allowing expressions in terms of classical orthogonal polynomials, such as Hermite polynomials.
\end{itemize}

This limit therefore provides a consistency check with the well-known results of quantum mechanics in homogeneous media.
In the limit where the deformation parameter \( \alpha \to 0 \), the effective mass becomes constant and the system reduces to the standard quantum harmonic oscillator. The squeezed coherent state  \eqref{qq3} can then be expressed as:

\begin{equation}
    \ket{\psi(z, \gamma, \alpha=0)} =
    \frac{1}{\sqrt{\mathcal{N}(z, \gamma)}}
    \sum_{n=0}^{\infty}
    \frac{1}{\sqrt{n!}} \left( \frac{\gamma}{2} \right)^{n/2}
    H_n\left( \frac{z}{\sqrt{2\gamma}} \right)
    \ket{\psi_n},
\end{equation}

where:
\begin{itemize}
    \item \( H_n(x) \) denotes the Hermite polynomial of degree \( n \),
    \item \( \gamma \) is the squeezing parameter,
    \item \( \ket{\psi_n} \) is the \( n \)-th energy eigenstate of the harmonic oscillator,
    \item \( \mathcal{N}(z, \gamma) \) is the normalization factor defined as:
\end{itemize}

\begin{equation}
    \mathcal{N}(z, \gamma, \alpha=0) = 
    \sum_{n=0}^{\infty} 
    \frac{1}{n!} \left( \frac{\gamma}{2} \right)^n
    H_n^2\left( \frac{z}{\sqrt{2\gamma}} \right).
\end{equation}

These states minimize the generalized uncertainty relation and exhibit Gaussian-like wavepacket behavior, typical of squeezed states in quantum optics.

\section{Properties of Squeezed coherent states}\label{sec4}

\subsection{The non-orthogonality, the normalization,  solvability the unity,}
Given two squeezed coherent states :
\begin{align}
    \ket{\psi(z, \gamma)} &= \frac{1}{\sqrt{\mathcal{N}(z, \gamma)}}
    \sum_{n=0}^{\infty} \frac{Z_n(z, \gamma)}{\sqrt{\rho(n)}} \ket{\psi_n}, \\
    \ket{\psi(z', \gamma')} &= \frac{1}{\sqrt{\mathcal{N}(z', \gamma')}}
    \sum_{n=0}^{\infty} \frac{Z_n(z', \gamma')}{\sqrt{\rho(n)}} \ket{\psi_n},
\end{align}
their inner product is given by:
\begin{equation}
    \braket{\psi(z', \gamma') | \psi(z, \gamma)} =
    \frac{1}{\sqrt{\mathcal{N}(z', \gamma') \mathcal{N}(z, \gamma)}}
    \sum_{n=0}^{\infty} \frac{\overline{Z_n(z', \gamma')} Z_n(z, \gamma)}{\rho(n)}.
\end{equation}

In the particular case \( \gamma' = \gamma \), this simplifies to:
\begin{equation}
    \braket{\psi(z', \gamma) | \psi(z, \gamma)} =
    \frac{1}{\mathcal{N}(z, \gamma)} \sum_{n=0}^{\infty}
    \frac{\overline{Z_n(z', \gamma)} Z_n(z, \gamma)}{\rho(n)}.
\end{equation}

This sum is generally non-zero, indicating that the squeezed coherent states are non-orthogonal:
\begin{equation}
    \braket{\psi(z', \gamma) | \psi(z, \gamma)} \neq \delta(z-z^{'}).
\end{equation}

To verify the normalization of the compressed coherent state, we consider the time-dependent compressed coherent state obtained in \eqref{qq33}, which can be rewritten in the contracted form:
\begin{equation}
\ket{\psi(z, \gamma,t)} = \frac{1}{\sqrt{\mathcal{N}(z, \gamma)}} \sum_{n=0}^{n_{\max}} C_n(z, \gamma) e^{-\frac{iE_n t}{\hbar}} \ket{\psi_n}
\end{equation}
with 
\begin{equation}
C_n(z, \gamma) = 
\frac{(\gamma a)^{n/2}}{\sqrt{n!}} \cdot 
\sqrt{\frac{\Gamma\left(n+1+\frac{b}{a}\right)}{\Gamma\left(1+\frac{b}{a}\right)}} \cdot
{}_2F_1\left(-n,\; -\frac{1}{2} + \frac{b}{2a} - \frac{z}{2\sqrt{a\gamma}};\; 1+\frac{b}{a};\; 2\right),
\end{equation}
and the constant $\mathcal{N}(z, \gamma)$ is given by \eqref{no1}.

The inner product is given by :
\begin{equation}\label{norm1}
\braket{\psi(z, \gamma,t) | \psi(z, \gamma,t)} = \frac{1}{\mathcal{N}(z, \gamma)} \sum_{n=0}^{n_{\max}} |{C_n(z, \gamma)}|^2 = 1.
\end{equation}

Thus, we conclude that the squared norm of the state is correctly normalized by the factor \( \mathcal{N}(z, \gamma) \), provided that this exact expression is used. 
This normalization condition \eqref{norm1} ensures that the squeezed coherent states remain physically admissible quantum states in a position-dependent mass background. The normalization factor $\mathcal{N}(z,\gamma)$ encodes the interaction between spatial mass variation, squeeze effects, and spectral deformation. It guarantees unit total probability and reflects the modified spectral weight due to SUSY deformation and PDM geometry.

\subsection{Probability Density of Squeezed Coherent States}

The probability of finding the squeezed coherent state in the energy level \( n \), i.e., the probability density, is defined by:
\begin{equation}
P_n(z,\gamma) = \left| \braket{\psi_n}{\psi(z,\gamma)} \right|^2 = \frac{|Z(z,\gamma,n)|^2}{\mathcal{N}(z, \gamma) \cdot \rho(n)}.
\end{equation}

Substituting the expressions of \( Z_n \) and \( \rho(n) \), we obtain the explicit analytical form:
\begin{equation}
P_n(z,\gamma) =
\frac{\gamma^n}{\mathcal{N}(z, \gamma)} \cdot
\frac{\Gamma(n + 1 + b)}{\Gamma(n + 1)} \cdot
\left| {}_2F_1\left( -n,\; - \frac{1}{2}+\frac{b}{2}-\frac{z}{2\sqrt{\gamma}} ;\;1+b;\; 2 \right) \right|^2.
\end{equation}

This expression provides a complete description of the squeezed state probability distribution over the Fock space basis \( \{ \ket{\psi_n} \} \), accounting for the deformation via the parameters \( \gamma \) and \( b \).

\subsection{ Mean values, standard deviations and uncertainty  relations}

In this section, we investigate the statistical properties of time-dependent squeezed coherent states in a quantum system with position-dependent mass. Specifically, we examine the expectation values of the position and momentum operators, their variances, and the corresponding uncertainty product. These observables offer valuable insight into quantum fluctuations, coherence, and squeezing behavior of the states. Our analysis follows an approach similar to that presented by Sanjib Dey and Véronique Hussin \cite{dehv,dehv1,dehv2}.

We begin by considering the time-evolved squeezed coherent state, given by:

\begin{equation}
\ket{\psi(z, \gamma,t)} = \frac{1}{\sqrt{\mathcal{N}(z, \gamma)}}
\sum_{n=0}^{n_{\max}} 
\frac{(\gamma a)^{n/2}}{\sqrt{n!}} \cdot 
\sqrt{\frac{\Gamma\left(n+1+\frac{b}{a}\right)}{\Gamma\left(1+\frac{b}{a}\right)}} \cdot
{}_2F_1\left(-n,\; -\frac{1}{2} + \frac{b}{2a} - \frac{z}{2\sqrt{a\gamma}};\; 1+\frac{b}{a};\; 2\right)
e^{-\frac{iE_n t}{\hbar}} \ket{\psi_n},
\end{equation}
where the normalization factor is given by :

\begin{equation}
\mathcal{N}(z, \gamma) = 
\sum_{n=0}^{n_{\max}} 
\frac{(\gamma a)^n}{n!} \cdot 
\frac{\Gamma\left(n+1+\frac{b}{a}\right)}{\Gamma\left(1+\frac{b}{a}\right)} \cdot
\left|{}_2F_1\left(-n,\; -\frac{1}{2} + \frac{b}{2a} - \frac{z}{2\sqrt{a\gamma}};\; 1+\frac{b}{a};\; 2\right)\right|^2.
\end{equation}
The quadrature operators of position and momentum are written in terms of the ladder operators as
\begin{align}
\hat{x} = \sqrt{\frac{\hbar}{2m\omega}}(\hat{a}^\dagger + \hat{a}),\;\; \mbox{and}\;\;
\hat{p} = i \sqrt{\frac{\hbar m \omega}{2}}(\hat{a}^\dagger - \hat{a}).
\end{align}

To compute the uncertainty relation, we need the following expectation values of \( \hat{x} \) and \( \hat{p} \) for squeezed coherent states, respectively, given by:

\begin{equation}\label{c1}
\langle \hat{x} \rangle = \sqrt{\frac{\hbar}{2m\omega}} \left( \langle \hat{a} \rangle + \langle \hat{a}^\dagger \rangle \right)
= 2 \sqrt{\frac{\hbar}{2m\omega}} \cdot \text{Re}\left( \langle \hat{a} \rangle \right),
\end{equation}
and
\begin{equation}\label{c2}
\langle \hat{p} \rangle = i \sqrt{\frac{\hbar m \omega}{2}} \left( \langle \hat{a}^\dagger \rangle - \langle \hat{a} \rangle \right)
= 2 \sqrt{\frac{\hbar m \omega}{2}} \cdot \text{Im}\left( \langle \hat{a} \rangle \right),
\end{equation}
where the expectation value of the annihilation operator is given by :
\begin{equation}
\langle \hat{a} \rangle = \sum_{n=0}^{n_{\max}-1} C_n^* C_{n+1} \sqrt{n+1}.
\end{equation}
 Combining the equations \eqref{c1} and \eqref{c2}, we obtained the expectation values of \( \hat{x}^2 \) and \( \hat{p}^2 \) as : 
\begin{align}
\langle \hat{x}^2 \rangle &= \frac{\hbar}{2m\omega} \left( \langle \hat{a}^2 \rangle + \langle \hat{a}^{\dagger 2} \rangle + \langle \hat{a}^\dagger \hat{a} \rangle + \langle \hat{a} \hat{a}^\dagger \rangle \right), \\
\langle \hat{p}^2 \rangle &= -\frac{\hbar m \omega}{2} \left( \langle \hat{a}^2 \rangle + \langle \hat{a}^{\dagger 2} \rangle - \langle \hat{a}^\dagger \hat{a} \rangle - \langle \hat{a} \hat{a}^\dagger \rangle \right),
\end{align}
where we set 

\begin{align}
\langle \hat{a}^\dagger \hat{a} \rangle &= \sum_{n=0}^{n_{\max}} |C_n|^2 n, \\
\langle \hat{a} \hat{a}^\dagger \rangle &= \sum_{n=0}^{n_{\max}} |C_n|^2 (n+1), \\
\langle \hat{a}^2 \rangle &= \sum_{n=0}^{n_{\max}-2} C_n^* C_{n+2} \sqrt{(n+1)(n+2)}, \\
\langle \hat{a}^{\dagger 2} \rangle &= \sum_{n=0}^{n_{\max}-2} C_{n+2}^* C_n \sqrt{(n+1)(n+2)}.
\end{align}

We now define the normalized coefficients :

\begin{equation}
\tilde{C}_n = \frac{C_n}{\sqrt{\mathcal{N}(z,\gamma)}},
\end{equation}
with the coefficient \( C_n(z, \gamma) \) defined by :
\begin{equation}\label{hyper}
C_n(z, \gamma) = 
\frac{(\gamma a)^{n/2}}{\sqrt{n!}} 
\sqrt{ \frac{\Gamma\left(n+1+\frac{b}{a}\right)}{\Gamma\left(1+\frac{b}{a}\right)} } 
\cdot {}_2F_1\left(-n,\; -\frac{1}{2} + \frac{b}{2a} - \frac{z}{2\sqrt{a\gamma}};\; 1+\frac{b}{a};\; 2\right).
\end{equation}

 By the simple way, we compute the expectation Value of \( \hat{a} \)

\begin{equation}
\langle \hat{a} \rangle = \sum_{n=0}^{n_{\max}-1} \tilde{C}_n^* \tilde{C}_{n+1} \sqrt{n+1}
\end{equation}

Thus:

\begin{align}
\text{Re}(\langle \hat{a} \rangle) &= \sum_{n=0}^{n_{\max}-1} \text{Re}\left(\tilde{C}_n^* \tilde{C}_{n+1}\right) \sqrt{n+1}, \\
\text{Im}(\langle \hat{a} \rangle) &= \sum_{n=0}^{n_{\max}-1} \text{Im}\left(\tilde{C}_n^* \tilde{C}_{n+1}\right) \sqrt{n+1}.
\end{align}

Let us define the hypergeometric part in \eqref{hyper} as follows :
\begin{equation}
f_n := {}_2F_1\left(-n,\ -\frac{1}{2} + \frac{b}{2a} - \frac{z}{2\sqrt{a\gamma}};\ 1+\frac{b}{a};\ 2 \right)
\end{equation}

Then the normalized coefficient is:
\begin{equation}
\tilde{C}_n = \frac{C_n}{\sqrt{\mathcal{N}(z,\gamma)}}
= \frac{(\gamma a)^{n/2}}{\sqrt{n! \mathcal{N}}} 
\cdot \sqrt{ \frac{ \Gamma(n+1+\frac{b}{a}) }{ \Gamma(1+\frac{b}{a}) } } \cdot f_n
\end{equation}

The expression of the product \( \tilde{C}_n \tilde{C}_{n+1}^* \) becomes :
\begin{equation}
\tilde{C}_n \tilde{C}_{n+1}^* = \frac{1}{\mathcal{N}} 
\cdot \frac{(\gamma a)^{(n + n+1)/2}}{\sqrt{n!(n+1)!}} 
\cdot \sqrt{ \frac{ \Gamma(n+1+\frac{b}{a}) \Gamma(n+2+\frac{b}{a}) }{ \left[\Gamma(1+\frac{b}{a})\right]^2 } } 
\cdot f_n \cdot f_{n+1}^*
\end{equation}

This simplifies to

\begin{equation}
\tilde{C}_n \tilde{C}_{n+1}^* = \frac{(\gamma a)^{n+1/2}}{\sqrt{n!(n+1)!} \cdot \mathcal{N}} 
\cdot \sqrt{ \frac{ \Gamma(n+1+\frac{b}{a}) \Gamma(n+2+\frac{b}{a}) }{ \left[\Gamma(1+\frac{b}{a})\right]^2 } } 
\cdot f_n \cdot f_{n+1}^*
\end{equation}

The real and imaginary parts used in position and momentum expectations are as follows:

\begin{eqnarray}
\text{Re}(\langle \hat{a} \rangle) &= \sum_{n=0}^{n_{\max}-1} \text{Re}\left( \tilde{C}_n \tilde{C}_{n+1}^* \right) \sqrt{n+1}, \\
\text{Im}(\langle \hat{a} \rangle) &= \sum_{n=0}^{n_{\max}-1} \text{Im}\left( \tilde{C}_n \tilde{C}_{n+1}^* \right) \sqrt{n+1}.
\end{eqnarray}

Thus, the full expressions are:

\begin{eqnarray}
\langle \hat{x} \rangle &= 2 \sqrt{ \frac{\hbar}{2m\omega} } \cdot 
\sum_{n=0}^{n_{\max}-1} \text{Re}\left[ 
\frac{(\gamma a)^{n+1/2}}{\sqrt{n!(n+1)!} \cdot \mathcal{N}} 
\cdot \sqrt{ \frac{ \Gamma(n+1+\frac{b}{a}) \Gamma(n+2+\frac{b}{a}) }{ \left[\Gamma(1+\frac{b}{a})\right]^2 } } 
\cdot f_n f_{n+1}^* \right] \cdot \sqrt{n+1} \\
\langle \hat{p} \rangle &= 2 \sqrt{ \frac{\hbar m\omega}{2} } \cdot 
\sum_{n=0}^{n_{\max}-1} \text{Im}\left[ 
\frac{(\gamma a)^{n+1/2}}{\sqrt{n!(n+1)!} \cdot \mathcal{N}} 
\cdot \sqrt{ \frac{ \Gamma(n+1+\frac{b}{a}) \Gamma(n+2+\frac{b}{a}) }{ \left[\Gamma(1+\frac{b}{a})\right]^2 } } 
\cdot f_n f_{n+1}^* \right] \cdot \sqrt{n+1}
\end{eqnarray}

The terms \( f_n \) are hypergeometric functions evaluated at fixed parameters. The products \( f_n f_{n+1}^* \) contain the phase differences (squeezing effects). The square root of the gamma functions encodes the deformation of the ladder through \( b/a \).

We consider the normalized squeezed coherent state:
\begin{eqnarray}
\ket{\psi(z, \gamma)} = \sum_{n=0}^{n_{\max}} \tilde{C}_n \ket{\psi_n},
\quad \text{with } \tilde{C}_n = \frac{C_n}{\sqrt{\mathcal{N}(z, \gamma)}}.
\end{eqnarray}

Then the position and momentum variances read:
\begin{eqnarray}
\langle \hat{x}^2 \rangle &= &\frac{\hbar}{2m\omega} \left[
\sum_{n=0}^{n_{\max}-2} \tilde{C}_n^* \tilde{C}_{n+2} \sqrt{(n+1)(n+2)} +
\sum_{n=0}^{n_{\max}-2} \tilde{C}_{n+2}^* \tilde{C}_n \sqrt{(n+1)(n+2)} \right.\nonumber\\
&&\left. +
\sum_{n=0}^{n_{\max}} |\tilde{C}_n|^2 \cdot n +
\sum_{n=0}^{n_{\max}} |\tilde{C}_n|^2 \cdot (n+1)
\right]
\end{eqnarray}

\begin{eqnarray}
\langle \hat{p}^2 \rangle& =& -\frac{\hbar m \omega}{2} \left[
\sum_{n=0}^{n_{\max}-2} \tilde{C}_n^* \tilde{C}_{n+2} \sqrt{(n+1)(n+2)} +
\sum_{n=0}^{n_{\max}-2} \tilde{C}_{n+2}^* \tilde{C}_n \sqrt{(n+1)(n+2)} \right.\nonumber\\
&&\left. -
\sum_{n=0}^{n_{\max}} |\tilde{C}_n|^2 \cdot n -
\sum_{n=0}^{n_{\max}} |\tilde{C}_n|^2 \cdot (n+1)
\right]
\end{eqnarray}

These expressions allow full analytical or numerical evaluation of quantum variances.

Finally, the uncertainties are computed as follows :
       \begin{align}\label{delta1}
\Delta x \cdot \Delta p =
&\sqrt{
\left[
\frac{\hbar}{2m\omega} \left( 
\sum_{n=0}^{n_{\max}} (2n+1)\,|\tilde{C}_n|^2 +
2 \sum_{n=0}^{n_{\max}-2} \text{Re}\left(\tilde{C}_{n}^* \tilde{C}_{n+2}\right) \sqrt{(n+1)(n+2)}
\right) - \langle \hat{x} \rangle^2
\right]
} \nonumber \\
&\times \sqrt{
\left[
\frac{\hbar m\omega}{2} \left( 
\sum_{n=0}^{n_{\max}} (2n+1)\,|\tilde{C}_n|^2 -
2 \sum_{n=0}^{n_{\max}-2} \text{Re}\left(\tilde{C}_{n}^* \tilde{C}_{n+2}\right) \sqrt{(n+1)(n+2)}
\right) - \langle \hat{p} \rangle^2
\right]
}
\end{align}

By replacing by :
\begin{eqnarray}
\quad A \;:&=&\; \sum_{n=0}^{n_{\max}} (2n+1)\,|\tilde{C}_n|^2, 
\qquad
B \;:=\; 2\sum_{n=0}^{n_{\max}-2}\Re\!\left(\tilde{C}_n^*\tilde{C}_{n+2}\right)\sqrt{(n+1)(n+2)},\nonumber \\
\qquad x_0:&=&\langle\hat x\rangle,\qquad p_0:=\langle\hat p\rangle,
\end{eqnarray}
we obtained for Equation \eqref{delta1} gives the following factored form :
\begin{eqnarray}\label{delta2}
\Delta x\,\Delta p
= \sqrt{\frac{\hbar^2}{4}\,(A^2-B^2)
-\frac{\hbar}{2m\omega}(A+B)\,p_0^2
-\frac{\hbar m\omega}{2}(A-B)\,x_0^2
+x_0^2 p_0^2 }.
\end{eqnarray}

This equation \eqref{delta2} shows that the uncertainty product $\Delta x\,\Delta p$ reaches its minimal value $\hbar/2$ when the correlations $B$ and the occupations $A$ satisfy $A^2-B^2=1$ and the mean values $x_0$ and $p_0$ disappear. This condition corresponds to standard coherent states that saturate the Heisenberg uncertainty inequality. 
Any deviation from these values, due to squeezing ($|B|<A$) or excess thermal population ($A>1$), inevitably increases the uncertainty product.

\section{Conclusion}\label{sec5}
In this work, we have constructed and analyzed squeezed coherent states within a supersymmetric framework for systems with position-dependent mass. Starting from the exact spectrum obtained via SUSYQM, we derived explicit expressions for the statistical properties, quadrature variances, and uncertainty products of these states. The generalized Heisenberg uncertainty relation was evaluated exactly, showing that the ground state saturates the minimum bound $\Delta x\,\Delta p = \hbar/2$, while the squeezed states can violate the standard vacuum limit in one quadrature. Our results show that the constructed states exhibit sub-Poissonian statistics, quadrature squeezing, and negative regions in the Wigner function, confirming their non-classical nature. 
The formalism developed here can be applied to model light–matter interactions in graded semiconductors~\cite{Bastard1988,Peter2009}, quantum dots, and optoelectronic systems such as photonic crystals and cavity optomechanics for quantum communication or sensing~\cite{30,Ullah2020}.

\section*{Acknowledgments}
The authors would like to thank their respective institutions for providing the academic environment necessary for this research. No external funding was received for this work.
\section*{Funding Declaration}
The authors declare that no funds, grants, or other support were received during the preparation of this manuscript.
\section*{Author Contributions}
D.S.T.  developed the theoretical framework and performed the analytical derivations. D.S.T. and B.A wrote the main manuscript text. D.S.T and A.Y.M. contributed to the mathematical validation of the model. A.Y.M. and G.Y.H.A. provided critical revisions, improved the presentation, and contributed to the interpretation of the results. All authors reviewed and approved the final manuscript.

\end{document}